\documentclass{article}

\usepackage{PRIMEarxiv}

\usepackage[utf8]{inputenc} 
\usepackage[T1]{fontenc}    
\usepackage{hyperref}       
\usepackage{url}            
\usepackage{booktabs}       
\usepackage{amsfonts}       
\usepackage{nicefrac}       
\usepackage{microtype}      
\usepackage{lipsum}
\usepackage{fancyhdr}       
\usepackage{graphicx}       
\usepackage{xcolor}
\graphicspath{{media/}}     
\usepackage{amsmath}
\usepackage{amssymb}    
\newcommand{\sgn}{\text{sign}}
\newcommand{\be}{\begin{equation}}
\newcommand{\ee}{\end{equation}}

\title{Modelling opinion misperception and the emergence of silence in online social system}

\author{
  Daniele Vilone$^{1,2}$, \ Eugenia Polizzi$^1$ \\
  \ \\
  $^1$ LABSS (Laboratory of Agent Based Social Simulation), \\ 
  Institute of Cognitive Science and Technology,
  National Research Council (CNR), Rome, Italy \\
  $^2$ Grupo Interdisciplinar de Sistemas Complejos (GISC), \\ Departamento de Matem\'aticas, Universidad Carlos III de Madrid, Spain \\ 
  \ \\
  \texttt{eugenia.polizzi@gmail.com; daniele.vilone@gmail.com} \\
}

\begin{document}
\maketitle

\ 

\begin{abstract}

In the last decades an increasing deal of research has investigated the phenomenon of opinion misperception in human communities and, more recently, in social media. Opinion misperception is the wrong evaluation by community's members of the real distribution of opinions or beliefs about a given topic. In this work we explore the mechanisms giving rise to  opinion misperception in social media groups, which are larger than physical ones and have peculiar topological features. By means of numerical simulations, we suggest that the structure of connections of such communities plays indeed a role in distorting the perception of the agents about others' beliefs, but it is essentially an indirect effect. Moreover, we show that the main ingredient that generates the misperception is a spiral of silence induced by few, well connected and charismatic agents, which rapidly drives the majority of individuals to stay silent without disclosing their true belief, leading minoritarian opinions to appear more widespread throughout the community.
\end{abstract}


\section{Introduction}

Adherence to group norms and group acceptance are among the most important factors shaping social behavior in
humans. The tendency of the individuals to conform to the behaviors, opinions and social norms of the others is an
innate trait of human beings, which sometimes can be in contradiction with the intimate beliefs of the individuals. This interplay between individual and social environment can have deep consequences on how humans communicate and transmit information ~\cite{HENRICH1998215} and importantly, on how opinions form and spread within groups~\cite{castellano2009statistical,sen2014sociophysics}.

People’s decisions are often guided by  what individuals anticipate the majority does or approves of~\cite{cialdini1998social, bicchieri2005grammar, ostrom2000collective}.
Importantly, these expectations does not need to be accurate to influence behavior:  biases may arise when inferring the beliefs of others and still guide decision-making. Extensive evidence from social psychology demonstrates the existence of such biases and their behavioral outcomes. For instance, individuals may misperceive others' beliefs due to false consensus bias, which is the tendency to attribute one's own belief to others~\cite{ross1977false,krueger1994truly}. Such a  bias can 
detrimentally affect public opinion processes by facilitating polarization and radicalization phenomena~\cite{wojcieszak2008false}.
Unpopular views may also become widespread due to pluralistic ignorance effects, a situation in which the majority of members privately disagree with a certain opinion or attitude, and yet publicly conform to it under the wrong belief that most others accept it~\cite{miller1987pluralistic,pre96}. Similar outcomes can also be induced by few "contrarians" loudly expressing their minoritarian opinions and by spiralling mechanisms that lead a disagreeing majority into silence ~\cite{noelle1974spiral}. 
Despite evidences of the effect of misperceptions in supporting unpopular behaviors and opinions in various real world settings~\cite{pre96,fields1976public,lambert2003pluralistic,sandstrom2013social,young2013delinquency,munsch2014pluralistic}, we still lack a clear understanding of how social biases emerge and impact opinion expression in online social systems.
Indeed, by enabling a few people to easily broadcast their opinions to millions of others~\cite{brennen2020types,krause2020fact}, while simultaneously increasing users' exposure to similar views ~\cite{nguyen2020echo,cinelli2021echo} social media may provide environments that systematically distort the inferences users make about the prevalence of certain opinions within their community, with critical consequences in terms of behavior at collective level~\cite{bessi2016social, lees2021understanding, moore2020exaggerated}.

Several findings support the role of vocal minorities in online debates \cite{gonzalez2013broadcasters, grinberg2019fake, mcclain2021behaviors} and of network structures in favoring the emergence of social biases. For example, a recent study analyzing the communication network structure of millions of COVID-related fake-news on Twitter provides evidence of an overexposure of a large group of passive users to the tweets of a few active users responsible for fabricating most of the contents available online. A limited understanding of the origin of the information may increase the chance that misinformation spreads due to users conforming to the "voice of few", wrongly perceived as representative of the "opinion of many"~\cite{castioni2021voice}. Observational data of online behavior however do not allow an in-depth analysis of the mechanisms responsible for the change in behavior, nor to identify the potential discrepancy between what individuals publicly declare and what they privately think. Yet, unfolding the complexity of such dynamics empirically is not trivial~\cite{axelrod1997dissemination,macy2002factors}. In this respect, agent-based modelling (ABM) may represent a useful approach, as it allows building simplified models of the dynamic under investigation along with an a-priori definition of key variables (e.g., agents’ decision-making rules) and system parameters (e.g., network structure) and to observe which macroscopic patterns can spontaneously emerge ~\cite{epstein1999agent}. Specifically, ABM consists in simulating social dynamics by means of virtual agents that interact among themselves following some established rules~\cite{conte2001sociology,conte2002agent,helbing2012agent}.
Recent effort in ABM has indeed been addressed the modelling of the dynamics of opinion expression, in particular the spiral of silence, by incorporating the potential for a disconnection between agents’ expressed and private opinions~\cite{ross2019social,sohn2016collective} and, in some cases, network heterogeneity~\cite{ross2019social, wu2015exploring,ma2021opinion}. While in principle these are adequate models to study opinion dynamics for specific scenarios (e.g., elections with major candidates, where binary options are equally likely to occur) they are less suited to examine the processes by which a small fraction of individuals holding  extreme positions (hereby "contrarians") can fuel opinion misperception in majority holding more moderate views and sway the public discourse toward their side. Simulations have also been used to explore whether perceptual social biases can emerge simply due to the geometry of the network, showing how simple topological effects 
can increase the perceived prevalence of few high-degree nodes in the eyes of their less connected neighbors: in the context of online systems this "illusion of majority" bias, rooted in the network structure, has indeed been suggested to act as an exogenous mechanism facilitating the spread of unpopular attitudes and behaviors (\cite{lerman2016majority}). 
Yet, this crucial point requires still to be empirically tested. Further simulation work that have modelled social biases as stemming from a wrong inference of others' beliefs have shown how "endogenous" sources of misperception (e.g., pluralistic ignorance mechanisms) play a key role in explaining how unpopular norms can spread in a system~\cite{merdes2017growing}.  
Models of this kind could also be applied to investigate how minorities can influence opinion expression in systems characterized by a heterogeneous (e.g., majority-minority) distribution of beliefs.

New models and theoretical approaches that account for both exogenous and endogenous determinants of opinion misperception are thus necessary to better understand the effect of social biases on the dynamics of opinion expression in complex communication networks. In this study we will thus address such a challenge by modelling communication among agents in a prototypical online community scenario: when a committed minority of users holding contrarians beliefs ( "contrarians") enters a population where some belief is commonly accepted though in diverse grade. For example, we can think about the exordia of the antivax movement, when few activists started to spread in internet their skepticism among communities which until have never questioned vaccinations~\cite{ler20,benoit2021anti}. 
By systematically varying Kc, the degree of nodes on which contrarians are placed, we aim to explore if (and so, how) minoritarian agents embedded in a network with social-media like features can ignite a change in the perceived public opinion and the repercussion of such misperception at the system level.

\ 

\section{Exogenous contributions to opinion misperception}
\label{TopMod}

In this section we explore whether a substantial distortion in the opinion distribution (as perceived by agents) can emerge mainly due to the complexity of the network, without resorting to the cognitive features of the agents and their interaction. More precisely, it has been pointed out that in complex networks the most connected nodes may be over-represented in the neighbourhood of other individuals, and this may induce agents to misperceive the actual opinion distribution: it is the so-called "frienship paradox"~\cite{caldarelli2007scale,caldarelli2012networks,feld1991your}, which states that in a heterogeneous topology the average degree of the neighbours of an agent is higher than the network's average degree itself.
Building on previous modelling work~\cite{lerman2016majority} we thus create a model where agents have minimal cognitive features and express their opinions in a heterogeneous communication network characterized by a clear majority-minority belief distribution. By manipulating the popularity of committed minoritarian agents we aim to evaluate the weight of topological heterogeneity in driving the dynamics of opinion misperception at macro-scale level.

\subsection{The Model}
\label{mod1def}

 We consider a system of $N$ agents arranged on a given network, defined by its adjacency matrix $\hat{M}_{ij}$ and characterized by its degree distribution $P(x)$~\cite{caldarelli2012networks}. As already stated, the internal structure of the agents and their interaction rules will be outlined in the simplest possible way. Specifically, each agent $i$ is defined as follows:

\begin{itemize}
    \item The belief, or private opinion, $b_i$, constant in time, which can assume one of two possible values,  $b_i=\pm1$;
    \item The strength of the opinion $\sigma_i\in(0,1]$, also constant in time, which specifies how strongly the agent holds its belief: when requested to publicly declare its opinion, the agent will answer its true belief with probability $\sigma_i$: this is the only  variable describing the internal state of the agents;
    \item The declared opinion $\omega_i$, which is what the agent tells the others about its own belief; it may or may not be equal to the private belief depending on the probability $\sigma_i$. At the initial stages of the dynamics, it may also assume the $0$ value besides $\pm1$ (see below for clarifications).
\end{itemize}

Once set the agents in the network and assigned the initial values of each variable (see below), the dynamics takes place. The elementary time-step of the dynamics is defined as follows:

\begin{itemize}
    \item An agent $i$ is selected at random;
    \item With probability $\sigma_i$, the agent will declare its private opinion, therefore $\omega_i=b_i$;
    \item With probability $1-\sigma_i$ the agent will declare the opinion expressed by  the majority of its own neighbours: 
$$
\omega_i=\sgn\left[\sum_{j\in\Gamma_i}\omega_j\right] \ , 
$$
   where $\Gamma_i$ is the set of the nearest neighbours of the agent $i$ (if the result is zero the expressed opinion will be $+1$ or $-1$ with equal probability);
   \item A time unit of the simulation will be given by $N$ of these elementary processes;
   \item In order to have enough statistics, all the results presented in the following are averaged over 10000 independent realizations.
\end{itemize}

Agents thus undergo a basic social influence process: the estimation of the majority opinion can only be based on what they can observe, i.e., the opinions expressed by their neighbours. Neither the global distribution of expressed opinion nor anyone else’s private beliefs is known. The adoption of such a heuristic is justified under a bounded rationality view of decision making, e.g., agents acting under limited information and limited capacity to make decisions. Therefore, the majoritarian opinion expressed by agents' neighbors is what agents can use to infer the "opinion climate", and will thus also correspond to agent's perception of the global distribution of private beliefs.

\noindent {\bf Topology:} we considered $N=5000$ agents on a Scale Free (SF) network with degree distribution with exponent $\lambda=2.2$, which characterizes with a good approximation the topological structure of many social media communities~\cite{caldarelli2007scale,fan2011online,varga2015scale,aparicio2015model}, and in particular on Twitter~\cite{omodei2015characterizing}. Therefore, we consider a network whose degree distribution follows the distribution below:
 
\be
P(k)\propto k^{-\lambda} \ , 
\label{deg_d}
\ee

\noindent where $\lambda=2.2$.
We generated such networks for our simulations by means of the Molloy-Reed algorithm~\cite{molloy1995critical}, so that for numerical reasons the minimum allowed degree is 2, the maximum $\lfloor\sqrt{N}\rfloor$~\cite{kryven2017general}.

\noindent {\bf State variables:} the first important variable describing the state of the system is $\langle\omega\rangle$, that is, the average declared opinion:

$$
\langle\omega\rangle=\langle\omega\rangle(t)\equiv\frac{\sum_{i=1}^N\omega_i(t)}{N} \ .
$$

Analogously, we consider the average private belief $b$, which is defined in the same way as the average declaration. For simplicity, in our model beliefs are not allowed to change over time. In order to evaluate the extent to which network features contribute to bias agents' perception of others' opinions we also include a variable measuring the opinion climate that an agent perceives in its immediate surrounding. In order to define such variable, we start from the neighbors' average declaration $\Delta_i$ of each agent:

$$
\Delta_i = \frac{\sum_{j\in\Gamma_i}\omega_j}{|\Gamma_i|}\ , 
$$

where $|\Gamma_i|$ represents the degree of the node $i$. Therefore, we define the average local opinion climate $r(t)$ as

$$
r(t)\equiv\langle\Delta_i(t)\rangle_i=\frac{\sum_{i=1}^N\Delta_i(t)}{N} \ . 
$$
Notice that all these quantities are defined in the real interval $[-1,+1]$.

As said, agents can only observe what is publicly expressed in their immediate surrounding, which may or may not be representative of the global distribution, and may or may not be aligned with the distribution of private beliefs. The potential discrepancy between the average opinion expressed around them (local opinion climate, which in our model is defined as the perception of the global belief distribution) and what agents declare on average overall (that is, the global opinion climate) will capture the contribution of topology to opinion misperception. 
Similarly, we can use the mismatch between the average opinion expressed around them and the true distribution of private, non observable beliefs to measure the full extent of misperception. 
In Table~\ref{tab:table} we provide a summary of the meaning of the main variables used to describe the state of the system. 

\begin{table}[htbp]
\begin{center}
\begin{tabular}{|c||c|}
\hline
Quantity & Definition \\
\hline
\hline
$\langle b\rangle$ & Average private opinion (belief)  \\
\hline
$\langle\omega\rangle$ & Average publicly declared opinion \\
\hline
$\Delta_i$ & Declared opinion averaged over agent' $i$ neighbours  \\
\hline
$r(t) = \langle\Delta_i\rangle$ & Local opinion climate $\equiv$ Average perception \\
\hline
$\langle b\rangle - \langle\omega\rangle$ & Private vs Public Mismatch (between beliefs and declarations) \\
\hline
$\langle b\rangle - r(t)$ & Misperception \\
\hline
$r(t) - \langle\omega\rangle$ & Topological contribution to misperception \\
\hline
\end{tabular}
\end{center}
\caption{\ List of the quantities utilized for the system's description.}
\label{tab:table}
\end{table}

\ 

\noindent {\bf Initial conditions:} 
The initial settings are chosen in order to mimic a prototypical social media environment with few committed users holding unpopular opinions ("contrarians") and a majority of other users holding more moderate views. 
We thus assign to contrarians a belief $b =-1$ and maximum strength, (i.e., $\sigma = 1$) that is, they always express opinions in line with their private belief. The behavior of these agents also recalls the so-called "hard cores" in the definition by Noelle-Neumann and Matthes \& colleagues~\cite{noelle1974spiral,matthes2010spiral}, that is, people who stick to their opinion regardless of the prevailing opinion climate.
Crucially, contrarians are selectively placed on nodes with degrees $K_c$, in order to highlight the role of the network's topology. In every other node with degree $k\neq K_c$ agents will hold belief $b=+1$ and strength $\sigma$ picked up from a uniform random distribution between 0 and 1. It follows that the distribution of contrarians will be given by the same $P(k)$ given in Eq.~(\ref{deg_d}), that is,
$P(K_c)\propto K_c^{-\lambda}$.
Specifically, when $K_c$ is high the network will be composed by less contrarians, but very well connected. Contrarily, when $K_c$ is low there will be relatively more abundant but less connected contrarians in the population.  
We will thus study how the model behave by systematically varying $K_c$, the degree of nodes (and so their "prestige") on which contrarians are placed. 
Once set the initial distribution $\{b_i\}$, the first time an agent is required to declare its opinion, if no one of its neighbours has already declared, with probability $1-\sigma_i$ the agent will declare a null opinion: $\omega_i=0$.

\subsection{Results}

\begin{figure}
  \centering
  \includegraphics[width=143mm]{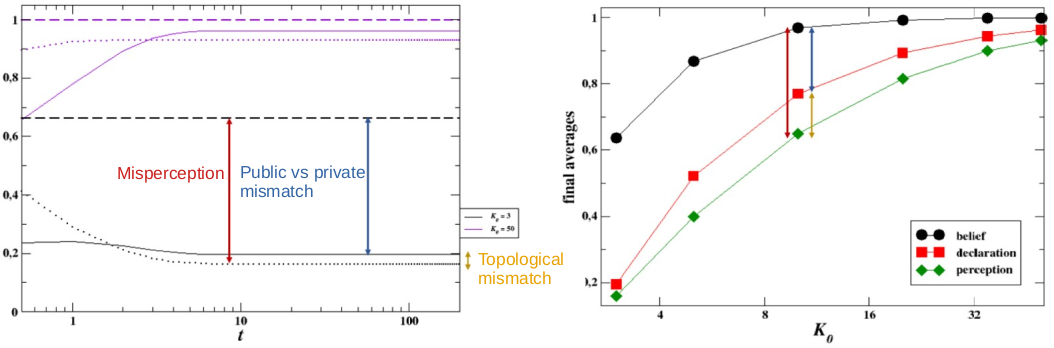}
  \caption{{\bf Left:} Time evolution of the average declared opinion $\langle\omega\rangle(t)$ (continuous lines), local opinion climate $r(t)$ (dotted lines), and average private belief $\langle b\rangle$ (dashed lines, constant in time) for two exemplifying scenarios: $K_c=3, \ 50$, in black and violet respectively. Scale-free network, size $N=5000$, $\lambda=2.2$, averages over 10000 independent realizations. We highlighted the contributions to misperception given by the topological features of the system (in yellow), i.e. network heterogeneity, and the endogenous one of the agents (in blue), i.e., the discrepancy between what agents believe and what publicly declare. {\bf Right: } Final values of the average declaration $\langle\omega_f\rangle$ (black line), average local opinion  $\langle r_f\rangle$ (green line), and average belief $\langle b\rangle$ (red lines) as functions of contrarians' degree $K_c$. Scale-free network, size $N=5000$, $\lambda=2.2$, averages over 10000 independent realizations. We highlighted again the contribution to misperception given by the topological features (yellow double arrow) and the one "endogenous" (i.e., due to the cognitive features of the individuals) of the agents, i.e., discrepancy between private vs publicly declared agents' opinion (blue double arrow).}

  \label{fig1}
\end{figure}

In Fig.~\ref{fig1} (left) we show the temporal evolution of $\langle\omega\rangle,\ r(t)$ and $\langle b\rangle$ for two different exemplifying values of contrarians' degree, $K_c=3,\ 50$ scenarios. For $K_c=3$  contrarians are relatively abundant ($\sim18\%$) but also poorly connected, while for the $K_c=50$ scenario contrarians are extremely few ($<0.001\%$) but also extremely powerful. In both scenarios, after an initial transient phase, both $\langle\omega\rangle$ and $r(t)$ reach a final stable value ($\langle b\rangle$ is by definition constant in time). Agents' perception of others opinion is always lower than the opinion expressed on overage by agents ($\langle\omega\rangle$), suggesting an over-exposure to contrarians' opinion in the local neighbourhood. However, such a distortion is rather small (about $5\%$). When taking into account the actual distribution of private beliefs $\langle b\rangle$, the discrepancy becomes considerably larger. 
This is easily observable in the $K_c=3$ scenario (the behavior for $K_c=50$ is qualitatively the same):  agents observe an average opinion around them of slightly less than 0.2, while the average private belief is around 0.67. 
Overall, the behavior of this model suggests that topology itself contributes only partially to the opinion misperception and that most of the phenomenon is due to the mismatch between what agents publicly declare and their private beliefs. Such a pattern implies that even when agents are modelled with minimal cognitive ingredients (i.e., strength of belief), the main source of bias appears to be endogenous (that is, mainly due to agents' behavior) rather than exogenous (that is, their position in the network). Fig.~\ref{fig1} (right) shows the final state of $\langle b\rangle,\ \langle\omega\rangle$ and $r$ as functions of the contrarians' degree. As can be observed, this same conclusion holds for every value of $K_c$ . With increasing $K_c$ all the curves tend to 1 because the size of the contrarians minority becomes drastically small, giving rise to a saturation effect. 

\ 

\section{Endogenous contributions to misperception}
\label{sec:others} 

The results obtained in the previous section show that topological effects alone are not enough to explain the overall misperception of the opinions held among the population, without resorting to more complex cognitive features of the agents. In order to better understand what drives the observed dynamics, we thus proceed by refining the characterization of the endogenous component of opinion misperception by acting on agents' interaction rules. 
Implementing complex agents' socio-cognitive features in ABM is a non-trivial task~\cite{flache2017models}, mainly because of the limits of encompassing the complexity of human cognition in a relatively simple algorithm. There are many ways to model the internal dynamics of agents, as for example refining models from Opinion Dynamics theory~\cite{giardini2015consensus}. 
A useful example comes from the study of Merdes~\cite{merdes2017growing} which investigated the role of belief misperception in the emergence of unpopular norms by means of simulations. In that model, agents have fixed, non observable, beliefs and are subject to social pressure from the surrounding neighborhood with regard to which behavior to adopt (e.g., to follow or not the norm by conforming to the most prevalent behavior). Agents' choice is driven by the effort to minimize the gap between their social expectations -- what agents believe most others do (or perceive so) -- and their own, potentially disagreeing, private belief. Such a characterization of agents' decision rule, along with the possibility for them to behave differently from their private beliefs creates the conditions for the emergence of misperception at macro-scale level, i.e., most agents behave in a way which is not aligned with what would be expected from the underlying distribution of private preferences in the population. Norms can be thought as a specific kind of opinion, and our differentiation between agents' private and publicly declared opinion (observable choice) allows us to adapt their formulation of social influence to describe the endogenous contribution to opinion misperception. Accordingly, in our model the social component will be formulated as the effort by the agents to minimize the gap between what they perceive is the majoritarian opinion around them and their own private – and potentially disagreeing -- opinion. As in~\cite{merdes2017growing}, such assumption should be able to capture human tendency to reduce as possible the deviations from others' behaviors, norms and opinions~\cite{bicchieri2005grammar}.
Crucially, people's perception of their social surrounding (whether in line or not with their private opinion) may not only affect which opinion to declare, but also, whether to declare at all. More precisely, the perceived disagreement with the opinion mostly declared in the agents' neighborhood is expected to limit agents likelihood to express their own private belief ~\cite{noelle1974spiral} and may thus further contribute to the emergence of opinion misperception in the system. Accordingly, we incorporate this insight by adding the possibility for agents to stay silent.

\subsection{The Model}

In order to describe the above processes mathematically, a specific {\it goal function} $G_i$ is introduced for every agent. This function gives the payoff gained by the agent according to its own and neighbors' state, so that the payoff increases when the state of the agent is closer to the others', but decreases when the agent is forced to change its own previous state. In other words, agents tend to minimize the disagreement with their neighbours' opinions as well as the cost to act or declare in contrast with ones' own private belief. More precisely, we assume that agents have a private opinion $b_i=\pm1$, constant in time, and an expressed opinion $\omega_i=0,\pm1$, where $\omega_i=0$ critically adds a preference for staying silent\footnote{In the first model, the option $\omega_i=0$ can be adopted only initially, when the agent does not declare its belief but no neighbor has still declared anything: therefore, it is not meant to represent agents' choice to stay silent.}. Again, to each agent $i$ is attached a belief's strength $\sigma_i$, which can assume a real value from 0 (no strength at all), to 1 (maximum of confidence in own belief). The dynamics in this case proceeds as follows: 

\begin{itemize}
    \item An agent $i$ is selected at random;
    \item To decide what to explicitly declare or whether to remain silent, the agent $i$ maximizes the {\it goal function} $G(\omega_i)$ defined as
    \be
    G(\omega_i) \equiv -\left |\frac{\sigma_ib_i+\sum_{j\in\Gamma_i}\mathcal{A}_i^j\ \omega_j}{|\Gamma_i|+1}-\omega_i \right| \ ,  
    \label{goal_f}
    \ee
    where $|\Gamma_i|\equiv\sum_{j\in\Gamma_i}|\omega_j|$, $\sigma_i$ is the strength of the belief of player $i$, and the function $\mathcal{A}_i^j$ is the "prestige" of the agent $j$ with respect to agent $i$ itself: we set $\mathcal{A}_i^j=\frac{\deg(j)}{\deg(i)}$ to incorporate the fact that more charismatic individuals tend to get more connections with others than less charismatic ones~\cite{pastor2002network,tur2021effect};
    \item Once found the value $\bar{\omega}_i$ which maximizes $G(\omega_i)$, such $\bar{\omega}_i$ will be the declaration choice (opinion expressed) by $i$ in the next interaction (if $\bar{\omega}_i=0$, agent $i$ stays silent).
    \item Notice that one could equivalently define the cost function $C(\omega_i)\equiv-G(\omega_i)$, and proceed in the same way with the only difference being that $C$ has to be minimized.
\end{itemize}

\ 

\noindent {\bf Topology:} As for the first model, we analyze Scale-Free networks with degree distribution $P_{sf}(k)\propto k^{-\lambda}$ with $\lambda=2.2$, as in Eq.~(\ref{deg_d}), generated again by means of a Molloy-Reed algorithm (see Sec.~\ref{TopMod}); we also accomplished some simulations with $\lambda=1.5$ in order to test the effect of increasing heterogeneity.
 
\noindent {\bf State variables: } To quantitatively characterize the model's behavior, we use the same variables defined in Subsec.~\ref{mod1def}, that is, the average declared opinion $\langle\omega\rangle$ and the average perception $r$ through the local opinion perception $\Delta_i$.

\noindent {\bf Initial conditions:} as for the first model, we consider systems with most agents having $b_i=1$ and few agents with negative opinions put on nodes of degree $K_c$; we set $\sigma_i=1$ for the latter agents, and uniformly distribute it in the interval $[0,1]$ for the former ones. Finally, we always assume $\omega_i=0\ \forall i$ at the very beginning of the dynamics.

As in the previous section, we will study how the model behaves by systematically varying $K_c$, the degree of nodes on which contrarians are placed, and all the results presented in the following section are  averaged over thousands independent realizations. 

\ 

\subsection{Results}

\begin{figure}[ht]
  \centering
  \includegraphics[width=77mm]{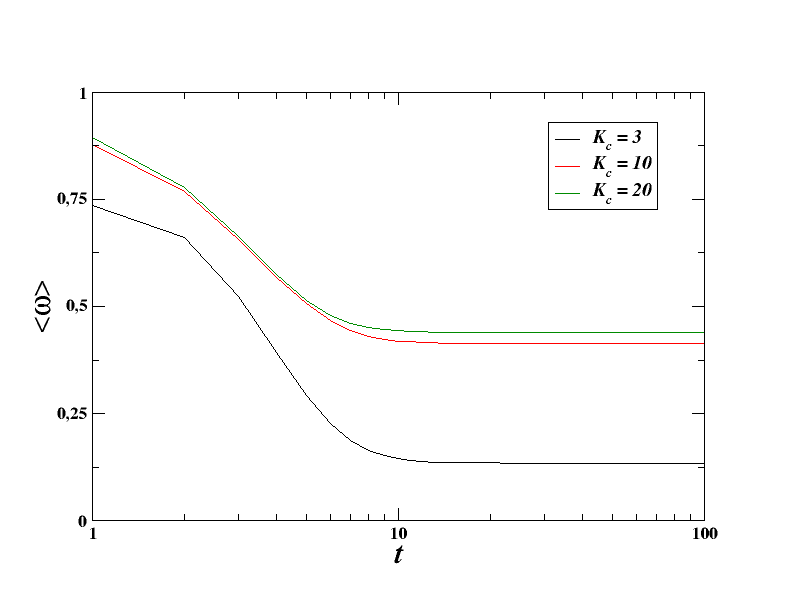} 
  \includegraphics[width=77mm]{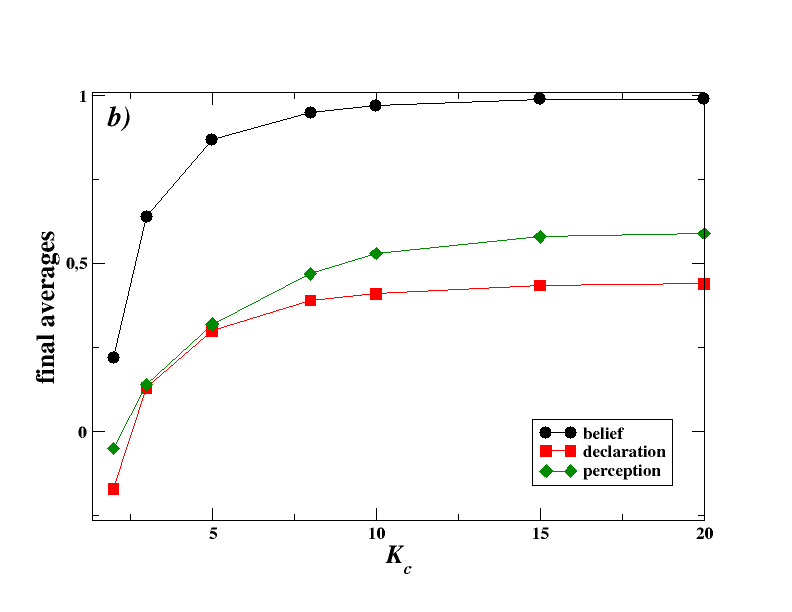}
  \caption{{\bf Left: }Temporal evolution of the average declaration for systems on SF networks with $\lambda=2.2$, and different values of $K_c$. System size $N=5000$, averages over 5000 independent realizations. {\bf Right: }behavior of the final average belief,  local perception and declaration as functions of the contrarians' degree $K_c$ for a highly heterogeneous SF networks,  $\lambda=2.2$. System size $N=5000$, results averaged over 5000 independent realizations.}
  \label{mod2_time}
\end{figure}

In Fig.~\ref{mod2_time} (left) we show the temporal evolution of agents' average declaration $\langle\omega\rangle$ for systems on SF networks ($\lambda=2.2,\ N=5000$), for three different exemplifying $K_c$ scenarios. Again, increasing $K_c$ values correspond to systems with fewer but better connected (and more "charismatic") contrarians. 

The average opinion expressed by agents is quite close to 1 at the initial stages of the dynamics (despite the fact that $\omega_i(t=0)\equiv0\ \forall i$, the agents declare their true belief the first time they express themselves), but decreases up to a relatively lower final value with similar qualitative behavior for different contrarians' degree. Also, the final declaration values are higher with increasing contrarians degree value, as the overall size of contrarians in the systems becomes drastically small (e.g., at  $K_c$ =20 only 3 out of 1000 agents are contrarians), but with a saturation effect for larger $K_c$. 

In Fig.~\ref{mod2_time} (right) we show the final values of the average belief,  local opinion climate and the actual opinion declaration, as a function of contrarians' degree $K_c$. The average belief value (which is constant in time) increases with $K_c$ because the number of contrarians decreases according to the increase in their degree. However, the average opinion declared by agents, both as it is perceived locally as well as it is declared globally, is always lower than the belief value in the population, and a stable plateau is reached with increasing $K_c$ values. This pattern is very different from the results of first model, where the average declaration value approached agents' belief values as $K_c$ increases (see Fig.~\ref{fig1} (right)). Such a finding suggests that, even when being of a negligible number,  well-connected contrarians are able to sway the global opinion towards their side, despite the fact that almost every agent holds a disagreeing private belief. In order to better understand the observed pattern, we look at how the proportions of agents choosing to declare the majoritarian or minoritarian opinion (or to stay silent) change according to $K_c$.

\begin{figure}[ht]
  \centering
  \includegraphics[width=91mm]{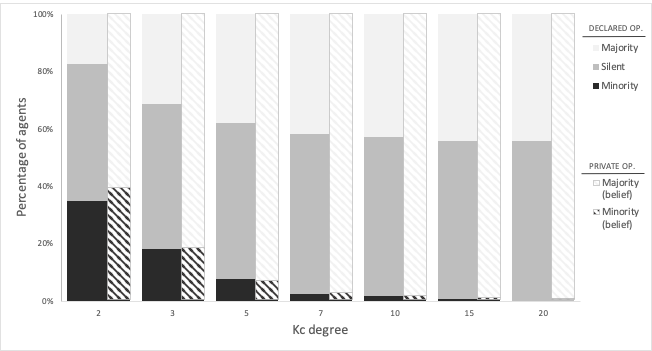}
  \caption{Frequency of the declared opinions, perception and beliefs in the system final state for different degree $K_c$ of the initial contrarians, distinguished among the different components of the population (majority, minority, silenced) are shown. Scale-free networks, size $N=5000$, $\lambda=2.2$.}
  \label{fig04b}
\end{figure}

\begin{figure}[ht]
  \centering
  \includegraphics[width=91mm]{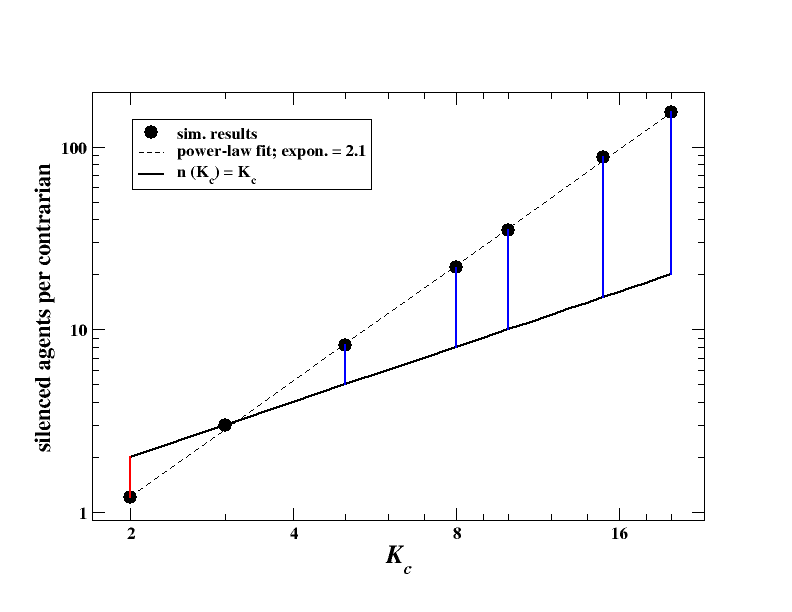} 
  \caption{{\bf Black circles:} number of silenced agents per contrarian as a function of the contrarians' degree $K_c$. {\bf Dashed line: } power-law fit $\sim K_c^{\nu},\ \nu\simeq 2.1$ (superlinear behavior). {\bf Red/blue lines:} difference between the number of agents silenced per contrarian and the actual number of contrarians' nearest neighbors. For $K_c=2$ the line is red because each contrarian drives less agents to silence than its number of nearest neighbors. {\bf System parameters:} SF network with exponent $\lambda=2.2$, system size $N=5000$; averages over 5000 independent realizations.}
  \label{cont_cont}
\end{figure}

Fig.~\ref{fig04b} shows the proportion of agents based on their final declaration choices $\omega$ at the end of dynamics for several values of $K_c$ from to 20 as compared to the fixed distribution of their private beliefs. The figure suggests that the shift towards the contrarian opinion is not due to majoritarian agents publicly aligning to contrarians' views, but to majority's choice of staying silent. Indeed, silence becomes the most effective solution for agents to minimize the cost of having a different belief from their (even though misperceived) immediate social surrounding without the need to deviate from their inner views. From a dynamical point of view, the possibility to avoid explicit declarations induces a cascade where agents surrounded by neighbours declaring mostly the opposite belief or being silent end up staying silent themselves, driving others to stay silent on their turn, and so on. Such a dynamics leads to a final state where the average declared opinion is much more skewed toward the opinion of contrarians compared to the average group’s private beliefs.
Interestingly, around $55\div60\%$ of subjects (almost everyone of which holding positive beliefs) end up being silent, regardless of contrarians' degree. In order to shed light on this effect, we compute the number of silenced agents per contrarian and plot it as a function of contrarians' degree. 
Fig~\ref{cont_cont} shows a superlinear relationship between the contrarians' degree value and the amount of majoritarian agents silenced. Specifically, when $K_c$ is low, e.g., 2, the size of the minority is large (around $35\%$), but every minoritarian agent drives less than 2 agents to silence. This suggests that for low $K_c$ silence mostly arises due to short-range dynamics, e.g., many contrarians directly influencing their immediate neighbors, and very few others. 
At $K_c=20$, the minority is negligible ($<0.1\%$), but for every minoritarian agent there are $\sim150$ majoritarian agents that get silenced, much more than their neighbours. Such a pattern suggests the presence of long-range dynamics, as silence spreads even beyond contrarians' direct connections. 
Therefore, for high  $K_c$ the increasing prestige of well-connected contrarians balances the reduction in their number, inducing the saturation effect that stabilizes the global declaration (and perception of local, too) toward lower values. Such a saturation effect is also reflected in the exponent of the superlinear relationship between contrarians and silenced agents, which is very close to the network $\lambda$ value: since the overall fraction of silent individuals remains approximately constant while the number of contrarians decrease as $K_c^{-\lambda}$, then the amount of the former versus the latter must increase with an exponent very close to $\lambda$ itself.

\begin{figure}[h]
  \centering
  \includegraphics[width=91mm]{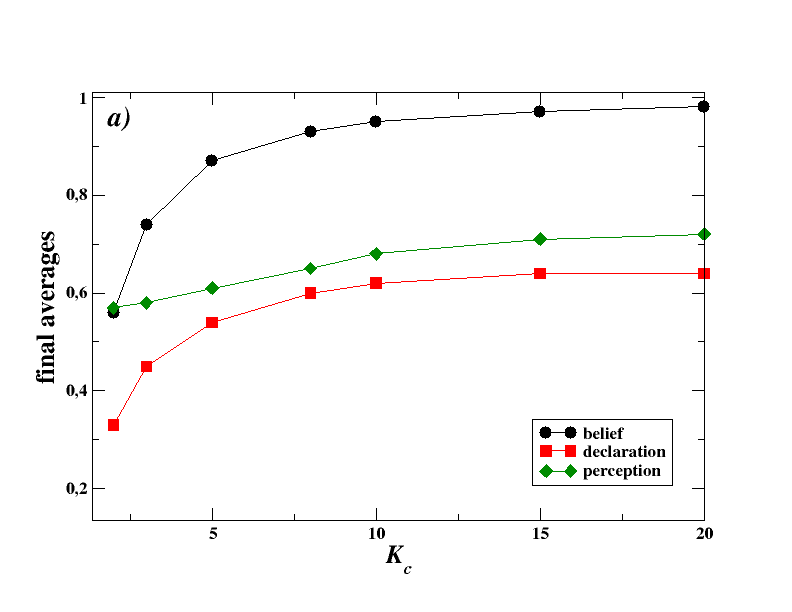}
  \caption{Behavior of the final average belief, local opinion and declaration  as functions of the contrarians' degree $K_c$ for a highly heterogeneous SF networks,  $\lambda=1.5$. System size $N=5000$, results averaged over 5000 independent realizations.} 
  \label{finconf_01_bis}
\end{figure}

As in the first model, the direct effect of topology in distorting agents' perception of others' opinion is again rather small. Interestingly, and in contrast to the former, the opinion that users can observe locally is actually closer to the true value of private beliefs as compared to the opinion declared on average by agents. Such an effect is possibly due to the stochasticity of the model, e.g., by few agents holding positive beliefs randomly ending on nodes with degree values even higher than the ones of contrarians. Such agents are thus able to limit the "traction" of contrarians with high prestige, whose effect is therefore top-down. Such a finding suggests that the direct topological effect on opinion misperception is, besides small, also non-universal.
Finally, 
we replicated the model on a Scale-free network with $\lambda=1.5$, corresponding to a more heterogeneous system than a Twitter-like one. From a qualitative point of view, a similar behavior can be observed (see Fig.~\ref{finconf_01_bis}), confirming that the results obtained are stable in the framework of SF networks.

\ 

\section{Discussion}

There is increasing evidence that social media communication features increase the potential for committed minorities to reach a much larger audience than in offline contexts, and to easily give rise to viral  phenomena~\cite{guerini2011exploring,feroz2014virality}. While in principle such effect may not be deleterious per se, an over-representation of otherwise minority opinions may become problematic if it increases the risk of inflating the support for socially questionable or even harmful views~\cite{castioni2021voice, ross2019social}. 
This study thus aims at better understanding how opinion misperception emerges as a results of agents dynamically interacting in networks with social-media like features. Specifically, we focused on evaluating the contribution of both exogenous (e.g., the structure of connections) and endogenous (e.g., agents' socio-cognitive features) mechanisms allowing committed minorities to be over-represented in the eyes of other users and the consequences of such distortion on the dynamics of opinion expression at collective level. 
Topological contribution was analyzed by selectively placing contrarians on nodes with degree $K_c$ and by running simulations for different degree $K_c$ values. Simulation results show that local network effects (in particular, the heterogeneity of the degree distribution) only marginally distort the perceived prevalence of minority opinions with respect to what agents declare on average. 
Besides being marginal, the effect of such topologically-induced bias in social perception also does not seem to be universal. Indeed, under some conditions the perceived prevalence of contrarians opinions in the local surrounding can even get closer to the actual distribution of agents’ private beliefs, thus partially correcting the effects of opinion misperception on the system. 
Previous modelling effort has shown how "illusion of majority" effects arising simply due to network features~\cite{lerman2016majority} can increase the perceived prevalence of minority opinions and act as network based explanation for the emergence of social biases. However, the actual impact of such distortions on collective behavior remains untested. By allowing agents to dynamically interact and by incorporating a differentiation between public vs private opinion we can explicitly test the effect of such distortions on public opinion.
Our results suggest that  behavior is driven by agents conforming under a wrong inference of contrarians' beliefs, which eventually shifts the global public opinion toward contrarians side.
Crucially, the result of the second model suggests that this effect is not due to a majority choosing to align with the contrarians minority view, but to a majority of agents choosing to stay silent. 
 
A spiral of silence emerges as a consequence of such individual choices.
Silence becomes thus a crucial element to explain the observed opinion dynamic patterns.
While similar mechanisms have already been noticed and discussed in literature~\cite{noelle1974spiral,scheufle2000twenty,matthes2015observing}, their existence in social media environments has been subject of debate~\cite{scheufle2000twenty,hayes2014self,katz2017six,matthes2018spiral}.
Observing spiralling processes and their effects in social media is difficult, mainly because only the views of those who express their opinions are evident there. Such a barrier can be lifted by means of computer simulations that allow to explicitly model the discrepancies between private and publicly expressed opinions.  
Although based on a simplistic description of opinion formation, our work suggests that the main feature of social media networks (i.e., the heterogeneity of connections) can promote social biases and contribute to amplify the voice of a motivated few by pushing into silence a disagreeing majority. 
Surprisingly, the silencing process emerges 
regardless of the network position or size of contrarians minority.
The present work underlines the importance of network characteristics when it comes to observing the dynamical process. Indeed, we show that when contrarians are located on low-degree nodes (with fewer direct connections), their lack in individual prestige is compensated by their larger number, which allow silencing process still to emerge at large scale. 
On the other hand, when placed on high-degree nodes, their numbers are few, but silence emerges due to their strong connections and high prestige, thus compensating for the significant reduction in their numbers. Consequently, these two effects balance each other, resulting in a consistently high fraction of silenced individuals. 
These findings align with the research conducted by Ross and colleagues~\cite{ross2019social}, who explored the impact of social bots (artificial, automated accounts impersonating humans) on diverting public discussions in social networks using a different model. Similarly to our findings, they show that the influence of these contrarians does not seem to be affected significantly by changes in their position or overall number. Notably, in both studies the percentage of silenced individuals reached levels around $50\%$ to $60\%$. Despite differences in the models and simulation algorithms employed, this agreement suggests that the presented findings may reflect an inherent characteristic of these types of social interactions.

As a consequence of the silencing process the proportion of agents expressing majority and minority opinions changes considerably with respect to the true distribution of private beliefs. However, in most of the investigated scenarios this process does not translate into a complete overturn of the majoritarian view. If we think of opinion formation and spread in social media, this suggests that in case of widely accepted views, few influential contrarians can weaken the strength of majoritarian position but this may not be enough to make them dominate public debate, as it has already been pointed out by Acerbi {\it et al.}~\cite{acerbi2019cultural,acerbi2022research}. 
In our model, the only case of inversion in majority-minority proportions is found when the minority exceeds the percentage of $30\%$. Interestingly, such a percentage is close to what indicated by Centola and Baronchelli~\cite{centola2018experimental} as being a tipping point 
beyond which a minority manages to spread globally into a population of agents.
Although our study does not allow (nor it was meant to) to investigate the effects of tipping points on the dynamics of opinion expression, further work could explore the link, i.e., whether opinion misperception can facilitate or hinder the emergence of social tipping points, and the role played by the network itself (e.g., the number and prestige of the agents involved).

\ 

In summary, our work contributes to the current literature on opinion dynamics in complex networks by exploring the determinants of opinion misperception in heterogeneous network systems. We tackle the issue by incorporating insights from socio-psychological literature to model the discrepancy between expressed and private opinions and the resulting effects at large-scale level. An open question is whether the results of our model can be translatable into “real” online scenarios: in order to address it, the model should be validated with empirical data. In related fields of research, experiments have shown that reducing the perceived prevalence of inappropriate behaviors in online fora (e.g., frequency of hate speech) significantly affects people willingness to engage in such actions by changing what people infer from the social context in terms of social acceptability (e.g., descriptive norms)~\cite{alvarez2018normative}. Similar experimental settings  could be designed to understand how changes in the (misperceived) descriptive norms may affect the likelihood that those with opposing views would speak out.
Data of this kind, along with observation of real networks and survey data could all provide valuable insights and empirical evidence to validate and expand our understanding of the processes behind opinion formation, misperception, and their dynamics at different levels. 
Finally, it has been shown that the effect of misperceptions can be alleviated by informing people about the real distribution of others' beliefs \cite{berkowitz2005overview}. If, as suggested here and elsewhere~\cite{castioni2021voice}, such biases occur in communication networks too, simulation models could be further used to test interventions to counteract the effects of such social biases, for example by testing whether revealing over-exposure to minority opinions can limit the emergence of silencing processes at system level. 
Multidisciplinary research combining experimental social science and computational methods can prove to be effective in enhancing our comprehension of the complex dynamics at play in online social interactions and help guide future research and interventions in this domain.

\ 

\subsubsection*{Acknowledgements}
This work was partially supported by project SERICS (PE00000014) under the MUR National Recovery and Resilience Plan funded by the European Union - NextGenerationEU and by the EU H2020 ICT48 project „Humane AI Net“ under contract n 952026. 

\newpage 

\bibliographystyle{unsrt}  
\bibliography{VP2022_main}

\end{document}